\documentclass[12pt]{iopart}  \usepackage{graphicx}
\usepackage{subfigure,cite} \usepackage{amsfonts,amssymb}
\newcommand{\tinytext}[1]{\mbox{\tiny{#1}}}
\newcommand{\text}[1]{\mbox{\scriptsize{#1}}}

\begin{document}

\title[Field-driven Polymer Translocation]{Pore-blockade Times for
Field-Driven Polymer Translocation}

\author{Henk Vocks$^{\dagger}$, Debabrata Panja$^*$, Gerard
  T. Barkema$^{\dagger,\ddagger}$ and Robin C. Ball$^{**}$}

\address{$\dagger$ Institute for Theoretical Physics, Universiteit
Utrecht, Leuvenlaan 4,\\  3584 CE Utrecht, The Netherlands

$^*$Institute for Theoretical Physics, Universiteit van Amsterdam,
Valckenierstraat 65,\\ 1018 XE Amsterdam, The Netherlands

$^{\ddagger}$Instituut-Lorentz, Universiteit Leiden, Niels Bohrweg 2,
2333 CA Leiden,\\ The Netherlands

$^{**}$Department of Physics, University of Warwick, Coventry CV4 7AL,
UK}

\begin{abstract}
We study pore blockade times for a translocating polymer of length
$N$, driven by a field $E$ across the pore in three dimensions. The
polymer performs Rouse dynamics, i.e., we consider polymer dynamics in
the absence of hydrodynamical interactions. We find that the typical
time the pore remains blocked during a translocation event scales as
$\sim N^{(1+2\nu)/(1+\nu)}/E$, where $\nu\simeq0.588$ is the Flory
exponent for the polymer. In line with our previous work, we show that
this scaling behavior stems from the polymer dynamics at the
immediate vicinity of the pore --- in particular, the memory effects
in the polymer chain tension imbalance across the pore. This result,
along with the numerical results by several other groups, violates the
lower bound $\sim N^{1+\nu}/E$ suggested earlier in the literature.
We discuss why this lower bound is incorrect and show, based on
conservation of energy, that the correct lower bound for the
pore-blockade time for field-driven translocation is given by $\eta
N^{2\nu}/E$, where $\eta$ is the viscosity of the medium surrounding
the polymer.
\end{abstract}


\maketitle

\section{Introduction\label{sec1}}
Molecular transport through cell membranes is an essential
mechanism in living organisms. Often, the molecules are too long, and
the pores in the membranes too narrow, to allow the molecules to pass
through as a single unit. In such circumstances, the molecules have to
deform themselves in order to squeeze --- i.e., translocate ---
themselves through the pores. DNA, RNA and proteins are such naturally
occurring long molecules \cite{drei,henry,akimaru,goerlich,schatz} in a
variety of biological processes. Translocation is also used in gene
therapy \cite{szabo,hanss}, and in delivery of drug molecules to their
activation sites \cite{tseng}. Consequently, the study of
translocation is an active field of research: as a cornerstone of many
biological processes, and also due to its relevance for practical
applications.

More recently, translocation has found itself at the forefront of
single-molecule-detection experiments \cite{nakane,expts,storm}, as
new developments in the design and fabrication of nanometer-sized pores
and etching methods may lead to cheaper and faster technology for the
analysis and detection of single macromolecules. In these experiments,
charged polymeric molecules, suspended in an electrolyte solution, are
initially located on one side of a membrane. The membrane is
impenetrable to the molecule except for a nanometer-sized
pore. Between the two different sides of the membrane, a DC voltage
difference is then applied, which drives the molecule through the
pore. When the molecule enters the pore, it affects the electrical
resistivity of the circuit, leading to a dip in the electric current
supplied by the voltage source. The magnitude and the duration
of these dips have proved to be very effective in determining the
size and the length of the molecule.  The usage of protein pores
(modified $\alpha$-haemolysin, mitochondrial ion channel, nucleic acid
binding/channel protein etc.) and the etching of specific DNA sequences
inside the pores \cite{szabo,proteinpore} have opened up promising new
avenues of fast, simple and cheap technology for single macromolecule
detection, analysis and characterization, perhaps even allowing DNA
sequencing at the nucleotide level.
\begin{figure}[!h]
\begin{center}
\includegraphics[width=0.5\linewidth,angle=0]{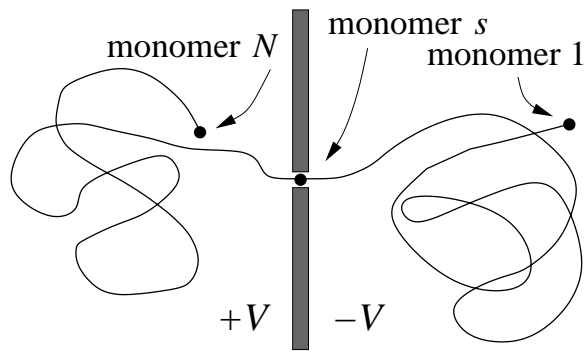}
\caption{Snapshot of a translocating polymer in a two-dimen\-sional
projection of our three-dimensional system. Across the pore of size
unity a voltage difference $2V$ is applied. The monomer located within
the pore is labeled $s$.
\label{geometry}}
\end{center}
\end{figure}

The subject of this paper is (charged) polymer translocation in three
dimensions through a narrow pore in an otherwise impenetrable membrane
placed at $z=0$, as the polymer is driven by a DC voltage across the
pore. Our interest is in the scaling behavior for the typical
pore-blockade time during a translocation event with polymer length
$N$. In practice, the electric field due to the applied voltage decays
rapidly with increasing distance from the pore, and for simplicity it is
often assumed that only those polymer segments residing within the
pore feel the driving force due to the field. For our theory and
simulations too, we consider a polymer which only experiences a force
acting on its monomers that reside in the pore, as illustrated in
Fig. \ref{geometry}.

To substantiate our theoretical analysis we use extensive Monte Carlo
simulations with a three-dimensional self-avoiding lattice polymer
model.  For the voltage difference across the pore we choose
\begin{equation} V(z)=\left \{\begin{array}{ll}+V & (z\leq-1) \\
                                    0  & (z=0) \\ -V &
                                    (z\geq1)\end{array}\right..
\label{potential}
\end{equation}
Thus, during translocation through the pore, the energy gained by each
monomer carrying a charge $q$, in dimensionless units, is given by
$\Delta U=2q V/k_BT$. From now on, favoring notational simplicity, we
choose both $q$ and $k_BT$ to be unity. Since we also choose
the lattice spacing to be unity in our simulations, the strength of
the electric field acting on each monomer within the pore is given by
$E=V$.

Details of the lattice polymer model used in this paper can be found
in Ref. \cite{modeldescription,anomlong}: the polymer moves through a
sequence of random single-monomer hops to neighboring lattice
sites. These hops can either be ``reptation''-moves, along the contour
of the polymer, or Rouse moves, in which the monomer jumps
``sideways'' and changes the contour. The definition of time used
throughout this paper is such that every monomer attempts a
``reptation''-move as well as a ``sideways''-move with rate
unity. There is no explicit solvent in our analysis, i.e., the polymer
performs Rouse dynamics.

Our conventions to study this problem, all throughout this paper, are
the following. We place the membrane at $z=0$. We fix the middle
monomer (monomer number $N/2$) of a polymer of total length $N$ at the
pore, apply the voltage as in Eq. (\ref{potential}) and thermalize the
polymer. At $t=0$ we release the polymer and let translocation
commence. We define the typical time when the polymer leaves the pore
as the dwell time $\tau_d$: it scales with $N$ in the same way as the
pore-blockade time in a full (field-driven) translocation event.

This problem has recently been studied in Ref. \cite{kantor}, in which
a lower bound $\propto N^{1+\nu}/E$ has been argued
for $\tau_d$. This lower bound was derived in the limit of unimpeded
polymer movement, i.e., for an infinite pore, or equivalently, in the
absence of the membrane. In Ref. \cite{kantor} the authors also
suggested that the dynamics of translocation is anomalous (see also
Ref. \cite{kantor2} in this context).

In the recent past, some of us have been investigating the microscopic
origin of the anomalous dynamics of translocation. We have set up a
theoretical formalism, {\it based on the microscopic dynamics of the
polymer}, and showed that the anomalous dynamics of translocation stem
from the polymer's memory effects, in the following
manner. Translocation proceeds via the exchange of monomers through
the pore: imagine a situation when a monomer from the left of the
membrane translocates to the right. This process increases the monomer
density in the right neighborhood of the pore, and simultaneously
reduces the monomer density in the left neighborhood of the pore.
The local enhancement in the monomer density on the right of the pore
\textit{takes a finite time to dissipate away from the membrane along
the backbone of the polymer\/} (similarly for replenishing monomer
density on the left neighborhood of the pore). The imbalance in the
monomer densities between the two local neighborhoods of the pore
during this time causes an enhanced chance of the
translocated monomer to return to the left of the membrane, thereby
giving rise to \textit{memory effects\/}. The ensuing analysis enabled
us to provide a proper microscopic theoretical basis for the anomalous
dynamics. Further theoretical analysis then led us to the conclusion
that in the case of unbiased translocation, i.e., when the polymer is
not subjected to an external force, the dwell time scales with length
as $\tau_d \sim N^{2+\nu}$ \cite{anom,anomlong,planar}, both in two
and three dimensions. Our approach based on the polymer's memory
effects also works beautifully for pulled translocation, during which
a force $F$ is applied at the head of the polymer: we have shown that
if $FN^\nu$ is sufficiently large, then the dwell time scales as
$\tau_d/N^{2+\nu} \sim (FN^\nu)^{-1}$ \cite{forced}. In this work, we
push ahead with the same formalism to demonstrate that it reveals
the physics of field-driven translocation too, thus providing
a {\it unified underlying theoretical basis\/} for translocation,
based on the theory of polymer dynamics.

\begin{table}[!h]
\begin{center}
\begin{tabular}{c|c|c}
$\quad\mbox{authors}\quad$& two dimensions& three
dimensions\tabularnewline \hline Kantor {\it et al.} \cite{kantor}&
$1.53\pm0.01$& $-$\tabularnewline \hline Luo {\it et al.} \cite{luo2}&
$1.72\pm0.06$ & $-$\tabularnewline \hline Cacciuto {\it et al.}
\cite{luijten}& $1.55\pm0.04$ & $-$\tabularnewline \hline Wei {\it et
al.} \cite{wei}& $-$& $1.27$\tabularnewline \hline  Milchev {\it et
al.} \cite{milchev}& $-$& $1.65\pm0.08$\tabularnewline \hline
Dubbeldam {\it et al.}  \cite{dubbeldam2}& $-$& $1.5$\tabularnewline
\hline
\end{tabular}
\caption{Existing numerical results on the exponent for the scaling of
$\tau_d$ with $N$ for field-driven translocation.  Note that the
proposed lower bound $1+\nu$ of Ref. \cite{kantor} is $1.75$ and
$1.59$ in two and three dimensions respectively.\label{table0}}
\end{center}
\end{table}
Returning to the lower bound for the scaling of the dwell time with
polymer length $N$ for field-driven translocation as proposed in Ref
\cite{kantor}, we note that subsequent numerical studies did not
immediately settle the scaling for $\tau_d$ with $N$, including the
one by the authors of Ref. \cite{kantor} themselves. In Table
\ref{table0} we present a summary of the existing numerical results on
the exponent for the scaling of $\tau_d$ with $N$ for field-driven
translocation. All results quoted are for self-avoiding polymers in
the absence of hydrodynamical interactions in the scaling limit.

More recently, this lack of consensus prompted three of us to
investigate the issue of field-driven translocation in two dimensions,
via a proxy problem, viz., polymer translocation in three dimensions
out of strong planar confinements \cite{planar}. We showed that the
actual lower bound for $\tau_d$ for field-driven translocation is
given by $\eta N^{2\nu}/E$, where $\eta$ is the viscosity of the
surrounding medium. This inequality is derived from the principle of
conservation of energy: it was shown in Ref. \cite{planar} that
although the presence of the memory effects suggests that the scaling
of $\tau_d$ could behave as $N^{(1+2\nu)/(1+\nu)}$, since
$(1+2\nu)/(1+\nu)<2\nu$ in two dimensions, conservation of energy
overrides the memory effects in the polymer --- high precision
simulation data suggested, in accordance with those of
Refs. \cite{kantor,luijten} that the actual scaling of $\tau_d$ for
field-driven translocation in two dimensions is given by $\tau_d\sim
N^{2\nu}$. In three dimensions $2\nu<(1+2\nu)/(1+\nu)$, implying that
in three dimensions $\tau_d\sim N^{(1+2\nu)/(1+\nu)}$, which is the
central result of this paper.

This paper is organized in the following manner. In Sec. \ref{sec2} we
derive the lower bound $N^{2\nu}$ for $\tau_d$ for field-driven
translocation. In Sec. \ref{sec3a} we discuss a method to measure the
polymer's chain tension at the pore. In Sec. \ref{sec3b} we analyze
the memory effects in the imbalance of the polymer's chain tension at
the pore. In Sec. \ref{sec4} we discuss the consequence of these memory
effects on the translocation velocity $v(t)$, and obtain the scaling
relation of $\tau_{d}$ with the polymer length $N$. We end this paper
with a discussion in Sec. \ref{sec5}.

\section{Lower bound for $\tau_d$ for field-driven
  translocation\label{sec2}}

As noted in Sec. \ref{sec1}, a lower bound for the dwell time
$\tau_d\sim N^{1+\nu}/E$ has been proposed in Ref. \cite{kantor}. The
underlying assumption behind this result is that, with or without an
applied field, the mobility of a polymer translocating through a
narrow pore in a membrane will not exceed that of a polymer in bulk
(i.e., in the absence of the membrane). This mobility is then obtained
under two more assumptions for the behavior of a polymer under a
driving field:
\begin{itemize}
\item[(i)] To mimic the field acting on a translocating polymer, the
field on the polymer in bulk has to act on a monomer whose position
along the backbone of the polymer changes continuously in time. As a
result, there is no incentive for the polymer to change its shape from
its bulk equilibrium shape, i.e., the polymer can still be described
by a blob with radius of gyration $\sim N^\nu$ in the appropriate
dimension.
\item[(ii)] The polymer's velocity is proportional to $DE$, where $E$ is the
applied field, and $D$ is the diffusion coefficient scaling as $D \sim
1/N$ for a Rouse polymer.
\end{itemize}
Of these two assumptions, note that (ii) is obtained as the steady
state solution of the equation of motion of a Rouse polymer, in bulk,
with uniform velocity and vanishing internal forces, see for instance
Ref. \cite{degennes}, Eq. VI.10. We have already witnessed in many
occasions
\cite{kantor,kantor2,klafter,dubbeldam,dubbeldam2,anom,anomlong,forced}
that the dynamics of translocation through a narrow pore is anomalous
(subdiffusive), as a consequence of the strong memory effects
discussed in the previous section, and also that these memory
effects are so strong that the velocity of translocation is not
constant in time \cite{forced,planar}. The anomalous dynamics and the
memory effects are crucial ingredients that question the validity of
the lower bound $N^{1+\nu}$ for $\tau_d$ for field-driven
translocation.

It is however possible to derive a lower bound for $\tau_d$ for
field-driven translocation, based on the principle of conservation of
energy. Consider a translocating polymer under an applied field $E$
which acts only at the pore. By definition, the $N$ monomers of the
polymer translocate through the pore in a time $\tau_d$. The total
work done by the field in this time $\tau_d$ is then given by
$EN$. During translocation, each monomer travels over a distance of
order $\sim R_g$, leading to an {\it average\/} monomer velocity
$v_m\sim R_g/\tau_d$. The rate of loss of energy due to the viscosity
$\eta$ of the surrounding medium per monomer is given by $\eta
v_m^2$. For a Rouse polymer, the frictional force on the entire
polymer is a sum of frictional forces on individual monomers, leading
to the total energy loss due to the viscosity of the surrounding
medium during the entire translocation event scaling as $\sim
N\tau_d\eta v_m^2=N\eta R^2_g/\tau_d$. This loss of energy must be
less than or equal to the total work $EN$ done by the field, which
yields us the inequality  $\tau_d\geq\eta R^2_g/E=\eta N^{2\nu}/E$
\cite{noteunbiased}.

\section{Memory effects in the chain tension perpendicular to the
  membrane \label{sec3}}

A translocating polymer can be thought of as two segments of
polymers tethered at the pore, while the segments are able to exchange
monomers between them through the pore. In Ref. \cite{anom} we
developed a theoretical method to relate the dynamics of translocation
to the imbalance of chain tension between these two segments across
the pore. The key idea behind this method is that the exchange of
monomers across the pore responds to $\phi(t)$, this imbalance of
chain tension; in its turn, $\phi(t)$ adjusts to $v(t)$, the transport
velocity of monomers across the pore. Here, $v(t)=\dot{s}(t)$ is the
rate of exchange of monomers from one side to the other.

The memory effects discussed in Sec. \ref{sec1} in terms of relaxation
of excess monomers (or the lack of monomers) in the immediate vicinity
of the pore translates immediately to that of the imbalance of the
chain tension across the pore --- local accumulation of excess
monomers reduce the chain tension, while local lack of monomers
enhance it. Quantitatively speaking, in the presence of memory
effects, the chain tension imbalance across the pore $\phi(t)$ and the
velocity of translocation $v(t)$ are  related by
\begin{eqnarray}
\phi(t)=\phi_{t=0}+\int_{0}^{t}dt'\mu(t-t')v(t')
\label{phi_t}
\end{eqnarray}
via the (field-dependent) memory kernel $\mu(t)$, which could be
thought of as time-dependent `impedance' of the system. Using the Laplace
transform, this relation could be inverted to obtain
$v(t)=\int_{0}^{t}dt' a(t-t')[\phi_{t=0}-\phi(t')]$, where $a(t)$ can
be thought of as the `admittance' of the system. In the Laplace
transform language, these are related to each other as
$\mu(k)=a^{-1}(k)$, where $k$ is the Laplace variable representing
inverse time \cite{anom,anomlong,forced,planar}.

\subsection{Chain tension perpendicular to the membrane\label{sec3a}}

Measuring chain tension directly is difficult. We therefore use a
method developed earlier  \cite{forced,planar} to monitor the chain
tension near the pore.

By definition, the chain tension imbalance $\phi(t)$ is the difference
of the chain tensions on the right and the left side of the pore:
$\phi(t)=\Phi_R(E,t)-\Phi_L(E,t)$. Both $\Phi_R(E,t)$ and
$\Phi_L(E,t)$ are functions of the applied electric field $E$ across
the pore. Note, from the applied potential (\ref{potential}), that the
field $E$ acts on the monomers at site $z=-1$ towards the pore, while
it acts on those at site $z=1$ away from the pore. Using the
convention that $E<0$ (resp. $E>0$) implies a field acting towards
(resp. away from) the membrane, we have
\begin{equation} \Phi(E,t=0)=\left\{\begin{array}{ll}\Phi_L(t=0) & (E<0) \\
                                    \Phi_R(t=0) &
                                    (E>0)\end{array}\right..
\label{phidef}
\end{equation}

Now consider a different problem, where one end of a polymer is
tethered to a fixed membrane, yet the number of monomers are allowed
to spontaneously enter or leave the tethered end, under the effect of
an electric field $E$.  Then, following the methodology described
in Refs. \cite{forced,planar}, we have
\begin{eqnarray}
\Phi(E,t=0)=k_BT\,\ln\frac{P_+}{P_-}\,,
\label{e3}
\end{eqnarray}
where $P_-$ (resp. $P_+$) is the probability that the left (or the
right) polymer segment has one monomer less (resp. one extra monomer).

\vspace{7mm}
\begin{figure}[ht]
\begin{center}
\includegraphics[width=0.45\linewidth,angle=270]{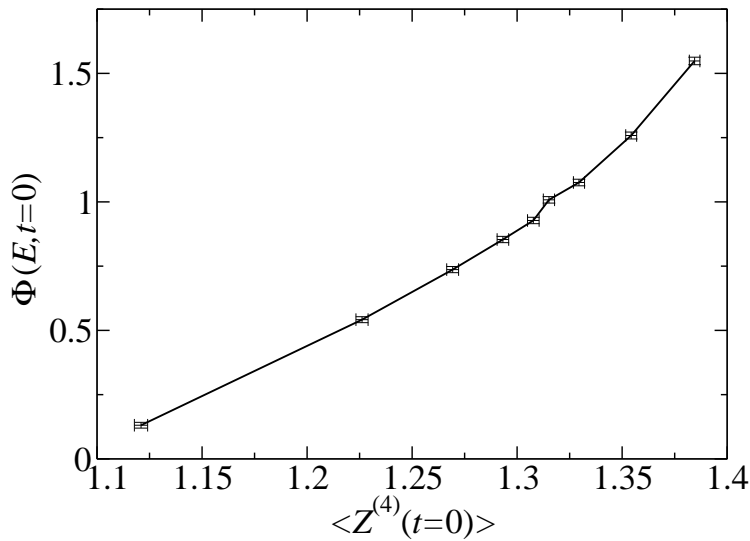}
\caption{$\langle Z^{(4)}(t=0)\rangle$ vs. $\Phi(E,t=0)$, for
$N/2=200$ and electric field values
$E=-0.5.-0.25,-0.1,-0.05,0,0.05,0.1,0.25$ and $0.5$ respectively. The
angular brackets for $\langle Z^{(4)}(t=0)\rangle$ indicate an average
over $32000$ polymer realizations, which are also used to obtain
$\Phi(E,t=0)$.\label{fig1}}
\end{center}
\end{figure}

Note that even for $E=0$, as already stressed in Ref. \cite{forced},
there is nonzero chain tension $\Phi_0$ at the pore, due to the
presence of the membrane. A polymer's free energy close to a membrane
is higher than its free energy in bulk. In other words, the membrane
repels the polymer, and as a result, for a polymer with one end
tethered to a membrane, the monomers close to the membrane are more
stretched than they would be in the bulk. 

For a translocating polymer Eq. (\ref{e3}) cannot be used, so to
compute $\Phi_R(t)$ and $\Phi_L(t)$ one needs a suitable proxy. In the
cases of unbiased translocation \cite{anom,anomlong}, translocation with
a pulling force \cite{forced} and translocation out of planar
confinements \cite{planar}, we have seen that the center-of-mass
distance of the first few, say $4$ to $5$ monomers from the membrane
provides an excellent proxy for $\Phi$. In this paper we follow the
same line. The average distance $\langle Z^{(4)}(t=0)\rangle$ is
plotted as a function of the chain tension $\Phi(E,t=0)$ for various
values of $E$ in Fig. \ref{fig1}. This figure shows that under an
applied field, $\Phi(E,t=0)$ is a reasonably linear function
well-proxied by $Z^{(4)}$.  The positive curvature seen in
Fig. \ref{fig1}, i.e., the deviation from linearity, is seen only for
$E>0$. We believe that this is partly due to the saturation of
$Z^{(4)}$. [By definition, in our lattice model the distance of the
center-of-mass of the first $4$ monomers from the membrane cannot
exceed $(1+2+3+4)/4=2.5$.]

\subsection{Memory effects in the chain tension \label{sec3b}} 

From Eq. (\ref{phi_t}), the behavior of the memory kernel
$\mu_R(t)$ for the polymer segment on the right side of the membrane
can be obtained with a sudden introduction of $p$ extra monomers at
the pore, corresponding to an impulse current $v(t)=p\delta(t)$. 
Physically, $v(t)=p\delta(t)$ with $p>0$ (resp. $p<0$) means that we
tether a polymer of length $N$ halfway through the pore at the pore at
$t\rightarrow -\infty$, let it thermalize till $t=0$, and then introduce
$p$ extra monomers at the tethered end of the right (resp. left)
segment at $t=0$. We then ask for the time-evolution of the mean
response $\langle\delta\Phi_R(t)\rangle$, where $\delta\Phi_R(t)$ is
the shift in chemical potential for the right segment of the polymer
at the pore. This means that for the translocation problem (with both
right and left segments), we would have
$\phi(t)=\delta\Phi_R(t)-\delta\Phi_L(t)$, where $\delta\Phi_L(t)$ is
the shift in chemical potential for the left segment at the pore due
to an opposite input current to it.

In earlier works \cite{anom,anomlong}, using 
$v(t)=p\delta(t)$ for a polymer of length $N$ tethered halfway at the
pore as described in the above paragraph, three of us showed that for
unbiased polymer translocation, i.e., for $E=0$, this mean response,
and hence $\mu(t)$ takes the form $\mu(t)\sim
t^{-\alpha}\exp[-t/\tau_{\tinytext{Rouse}}(N/2)]$  [note that for
$E=0$ there is a trivial symmetry between the right and the left
segment of the polymer, hence $\mu_R(t)=\mu_L(t)\equiv\mu(t)$]. 

\begin{figure}[!h]
\begin{center}
\includegraphics[width=0.45\linewidth,angle=270]{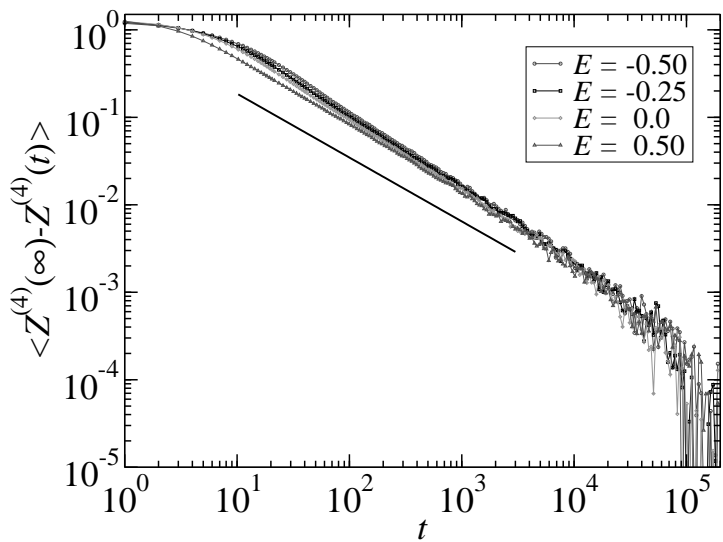}
\caption{Probing the memory kernels by $\langle
Z^{(4)}(\infty)-Z^{(4)}(t)\rangle$ following monomer injection at the
pore corresponding to $v(t)=p\delta(t)$, with $|p|=5$. Physically,
$v(t)=p\delta(t)$ with $p>0$ (resp. $p<0$) means that we tether a
polymer of length $N$ halfway at the pore at $t\rightarrow -\infty$, let
it thermalize till $t=0$, and then introduce $|p|$ extra monomers to the
right (resp. left) segment at the tether point at $t=0$. Following our
notation in Eq. (\ref{phidef}), the $E<0$ data (resp. $E>0$ data)
correspond to $\mu_L(t)$ [resp. $\mu_R(t)$].  The data presented
correspond to an average over $500,000$ polymer realizations, with
$N/2=200$. The steeper drop at longer times correspond to the
exponential decay $\exp[-t/\tau_{\tinytext{Rouse}}(N/2)]$. The solid
line corresponds to the power law $t^{-\frac{1+\nu}{1+2\nu}}\approx
t^{-0.73}$.
\label{fig2}}
\end{center}
\end{figure}

When the electric field is applied at the pore, and the same monomer
injection method is used to probe the memory kernels $\mu_R(t)$ and
$\mu_L(t)$, we expect the above arguments to hold again: since the
field is applied very locally at the base of the tethered polymer
segments, it does not destroy the broader structure of the
polymer. However, we do expect to see deviations from the
$t^{-\alpha}\exp[-t/\tau_{\tinytext{Rouse}}(N/2)]$ at short times.
Indeed, we have confirmed this picture --- for various field strengths
we tracked $\langle\delta\Phi_R(t)\rangle$ and
$\langle\delta\Phi_L(t)\rangle$ by measuring the distance of the
average centre-of-mass of the first $4$ monomers from the membrane,
$\langle Z^{(4)}(t)\rangle$, in response to the injection of extra
monomers near the pore at $t=0$. Specifically we consider the
equilibrated right and left segments of the polymer, each of length
$N/2=200$ (with  the middle monomer threaded at the pore), adding $5$
extra monomers at the tethered end of the right and the left segment
each at $t=0$, corresponding to $|p|=5$, bringing the length of each
segment up to $N/2+|p|$. Using the proxy $\langle Z^{(4)}(t)\rangle$ for
both segments we then track $\langle\delta\Phi_R(t)\rangle$ and
$\langle\delta\Phi_L(t)\rangle$, denoting them by values $E>0$ and
$E<0$ respectively in Fig. \ref{fig2}. The deviations from the
expected power-law $t^{-(1+\nu)/(1+2\nu)}$ at short times and the
$\exp[-t/\tau_{\tinytext{Rouse}}(N/2)]$ at long times makes the
precise identification of the power-law $t^{-(1+\nu)/(1+2\nu)}$
difficult. Nevertheless, there is an extended regime where this power
law can be identified reasonably clearly, yielding us
$\mu_R(t)=\mu_L(t)\equiv\mu(t)=t^{-(1+\nu)/(1+2\nu)}\exp[-t/\tau_{\tinytext{Rouse}}(N/2)]$.

\vspace{7mm}

\section{Scaling behavior of $\tau_d$ with $N$\label{sec4}}

The memory kernel we obtained in Sec. \ref{sec3} can be termed as the
``static memory kernel'', as it is obtained under the condition that
before the injection of the extra monomers both segments were
thermalized. When the applied field is not too strong, 
we can expect the static memory kernel to yield the scaling of
translocation velocity with time, in the following manner.

An inverse Laplace transform of Eq. (\ref{phi_t}) yields us
\begin{eqnarray}
v(k)=\frac{\phi_{t=0}}{k\mu(k)}\,-\,\frac{\phi(k)}{\mu(k)}\,,
\label{e6}
\end{eqnarray} 
where $k$ is the Laplace variable representing inverse time.
Thereafter, using the power-law part of $\mu(t)\sim
t^{-(1+\nu)/(1+2\nu)}$, i.e., $\mu(k)\sim k^{(1+\nu)/(1+2\nu)-1}$, and
Laplace-inverting Eq. (\ref{e6}), we get
\begin{eqnarray}
v(t)=\int_{0}^{t}\mathrm{d}t'\,(t-t')^{-\frac{1+3\nu}{1+2\nu}}\,\left[\phi_{t=0}-\phi(t')\right]\,.
\label{e7}
\end{eqnarray} 
If $\phi(t)$ goes to a constant $\neq\phi_{t=0}$, then Eq. (\ref{e7})
reduces to
\begin{eqnarray}
v(t)\sim t^{-\frac{\nu}{1+2\nu}},\,\,\mbox{i.e.,}\,
s(t)=N/2+\!\int_{0}^{t}\mathrm{d}t'\,v(t')\sim
t^{\frac{1+\nu}{1+2\nu}},
\label{e8}
\end{eqnarray} 
where $[s(t)-N/2]$ is the distance unthreaded after time $t$
\cite{note1}; the $N/2$ appears in Eq. (\ref{e8}) as $s(0)=N/2$.

\vspace{7mm}
\begin{figure}[!h]
\begin{center}
\includegraphics[width=0.45\linewidth,angle=270]{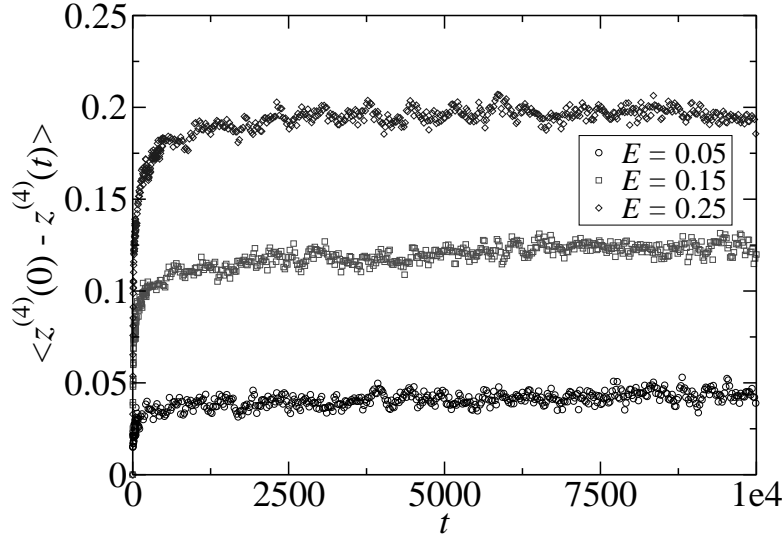}

\vspace{3mm}
\caption{Behaviour of $[\phi_{t=0}-\phi(t)]$ for $N=200$ as a function
of $t$, shown by means of the proxy variable
$<z^{(4)}(0)-z^{(4)}(t)>$.  demonstrating that $[\phi_{t=0}-\phi(t)]$
reduces to a constant very quickly: $E=0.05$ (circles), $0.15$
(squares), and $0.25$ (triangles). To generate these averages $16,000$
individual polymers were unthreaded for each value of $E$.\label{fig3}}
\end{center}
\end{figure}

In Fig. \ref{fig3} we show the behavior of
$\left[\phi_{t=0}-\phi(t)\right]$ by means of the proxy variable
$\langle z^{(4)}(0)-z^{(4)}(t)\rangle$ for $E=V=0.05$, $0.15$, and
$0.25$ respectively, where $z^{(4)}$ is the difference between the
$Z^{(4)}$ values of the right and left segment of the polymer, i.e.,
$z^{(4)}(t)=Z^{(4)}_R(t)-Z^{(4)}_L(t)$. Indeed the quantity
$\left[\phi_{t=0}-\phi(t)\right]$ approaches a constant rather
quickly. We also note that the relation between this constant and the
applied field $E$ is almost linear.

\vspace{7mm}
\begin{figure}[!h]
\begin{center}
\includegraphics[width=0.45\linewidth,angle=270]{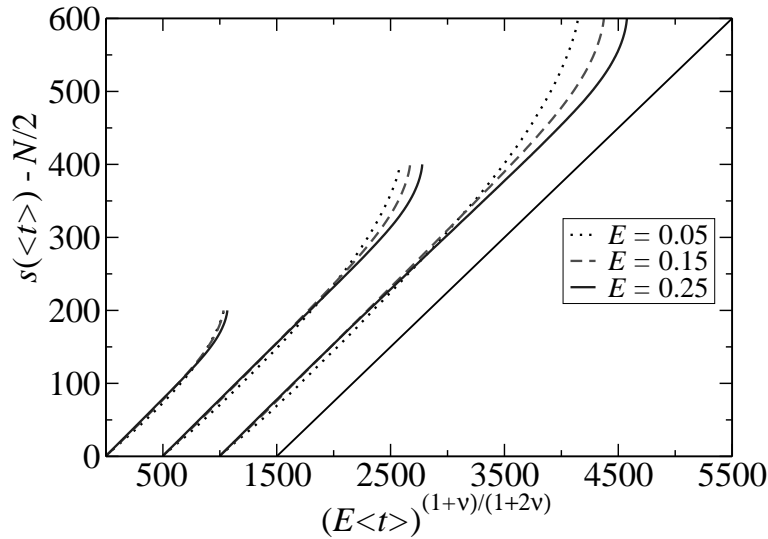}
\caption{The average time $\langle t\rangle$ to unthread distance $s$
for three different field strengths, for $N=400$ (average over
$16,000$ polymer realizations for each field), $N=800$ (average over
$16,000$ polymer realizations for each field), and $N=1,200$ ($5,000$
polymer realizations for $E=0.05$, and $7,500$ polymer realizations
each for $E=0.15$ and $E=0.25$). The data for $N=800$ correspond to
real time value, while the data for $N=400$ and $N=1,200$ have been
shifted by $\mp 500$ units along the x-axis for clarity. The solid
line has been added for a guide to the eye.
\label{fig4}}
\end{center}
\end{figure}

For strong fields, there is no {\it a priori\/} reason that the
dynamics can still be described by the static memory kernel instead of
a suitably replacing ``dynamic memory kernel'', but we find that the
scaling $s(t)\sim t^{(1+2\nu)/(1+\nu)}$ is obeyed for fairly strong
fields as well: in Fig. \ref{fig4} we plot the average time $\langle
t\rangle$ to unthread a distance $s$ to show this scaling. Note the
strong finite-size effects for the scaling behavior as shown by the
deviation from the $t^{(1+2\nu)/(1+\nu)}$ for larger values of
$s$. The presence of such strong finite-size effects indicates that
without the aid of $s(t)$ vs. $t$ curves, determining the scaling of
$\tau_d$ with $N$ will almost certainly lead to erroneous
identification of the scaling laws --- we believe that these
finite-size effects are responsible for the wide range of existing
numerical scaling results, as summarized in Table
\ref{table0}. Nevertheless, Fig. \ref{fig4} shows that these
finite-size effects do not increase linearly with $N$, leading us to
the scaling for $\tau_d$ as

\begin{eqnarray}
\tau_d \sim N^{(1+2\nu)/(1+\nu)}/E\,,
\label{finaleq}
\end{eqnarray}
which is obtained from the condition that $s(\tau_d)=N$. For
the above analysis to hold, the dwell time must be less than
$\tau_{\tinytext{Rouse}}$, which Eq. (\ref{finaleq}) confirms.  Note that
the $E$-dependence of Eq.  \ref{finaleq} is only numerically obtained
from Fig. \ref{fig4}.  Note also that the curves in Fig. \ref{fig4}
for $E=0.05$ tend to `sag' a bit.  We attribute this to our numerically
inspired definition of $s(\langle t\rangle)$, as the mean time to unthread
a distance $s$, as opposed to, e.g., the numerically less favorable
measure of distance $\langle s(t)\rangle$, i.e.,the monomer which
is most likely to reside in the pore at time $t$. At small fields,
the polymer has ample time for fluctuations, pushing the time of
first arrival up.  Numerically, for $E=0.15$ and $0.25$, the exponent
$\partial[\log{s\left(t\right)}]/\partial(\log{t})$ is found to be $0.73
\pm 0.02$, in agreement with the theoretical value $(1+\nu)/(1+2\nu)$.
The sagging and finite-size effects discussed above cause the apparent
exponent to be slightly larger, ranging from $0.74$ to $0.79$, for
$E=0.05$.

With decreasing field strength, especially in the range where the
thermal fluc\-tuations are comparable to the work done by the field
to translocate the entire polymer, given by $EN\simeq k_BT=1$, one
should obtain a crossover from the above scaling (\ref{finaleq})  to
$\tau_d\sim N^{2+\nu}$ for unbiased translocation
\cite{anom,anomlong}. This suggests that if $\tau_d/N^{2+\nu}$ is
plotted as a function of $EN$, then one should obtain a scaling
collapse; i.e., there exists a scaling function $f$ such that
$\tau_d=N^{2+\nu}f(EN)$. However, $EN$ as a scaling variable is
simply numerically inconsistent with Fig. \ref{fig4} and
Eq. (\ref{finaleq}). Instead $\tau_d=N^{2+\nu}f(E,N)$ is the proper
description of the situation, with $f(E,N)$ approaching a constant for
$E\rightarrow0,N\rightarrow\infty$ and $f(E,N)$ behaving as
$E^{-1}N^{-\nu-1/(1+\nu)}$ for $E\sim O(1)$ and
$N\rightarrow\infty$. Note that $E$ in this paragraph should be
interpreted as the dimensionless quantity $qV/(k_BT)$.

\vspace{7mm}
\begin{figure}[!h]
\begin{center}
\includegraphics[width=0.45\linewidth,angle=270]{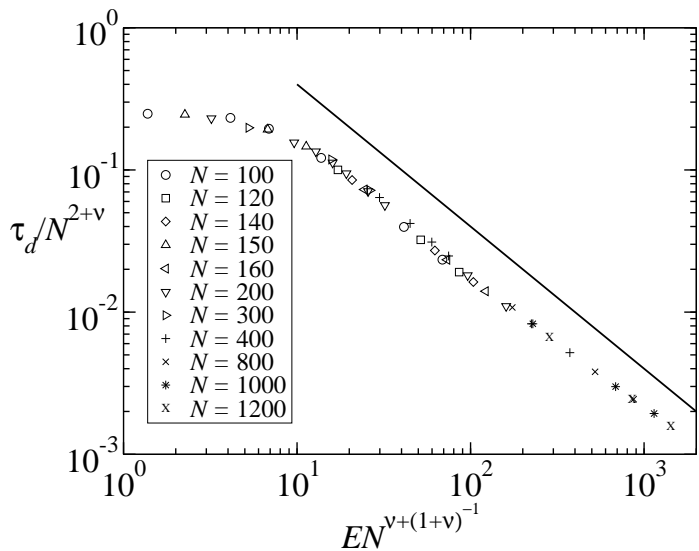}
\caption{Data collapse in terms of $x=EN^{\nu+1/(1+\nu)}$ and
$y=\tau_d/N^{2+\nu}$ for various fields. The mean unthreading times
$\tau_d$ have been obtained with at least $1,000$ polymers (up to
$32,000$ for smaller $N$ values)  for each $N$ and field strength $E$:
the statistical error bars are smaller than the size of the
symbols. The solid line $y\sim1/x$ for moderate field strengths in
support of Eq. (\ref{finaleq}) has been added for a guide to the eye.
\label{collapse}}
\end{center}
\end{figure}

To demonstrate the scaling behavior of Eq. (\ref{finaleq})
for $E\simeq O(1)$, we plot $\tau_d/N^{2+\nu}$ as a function of
$EN^{\nu+1/(1+\nu)}$ in Fig. \ref{collapse}. Keeping in mind that this
way of plotting the data does not necessarily yield a data collapse at
small but nonzero $E$, as discussed above, we also plot several data
points for small $E$, in order to demonstrate that for $E\rightarrow0$
our results in this paper are consistent with that of unbiased
translocation \cite{anom,anomlong,planar}.

\section{Discussion \label{sec5}}

In this paper, we studied polymer translocation in three dimensions
through a narrow pore in an otherwise impenetrable membrane, as the
polymer is driven by a field $E$ across the pore. The polymer performs
Rouse dynamics, i.e., we considered polymer dynamics in the absence of
hydrodynamical interactions. We found that the typical time the pore
remains blocked during a translocation event, for moderate field
strengths scales as $\sim N^{(1+2\nu)/(1+\nu)}/E$, where
$\nu\simeq0.588$ is the Flory exponent for the polymer. In line with
our previous works, we showed that this scaling behavior stems from
the polymer dynamics at the immediate vicinity of the pore --- in
particular, the memory effects in the polymer chain tension imbalance
across the pore \cite{anom,anomlong,forced,planar}. We also showed
that our results in this paper are consistent with that of unbiased
translocation \cite{anom,anomlong,planar} in the limit
$E\rightarrow0$.

The above results for finite $E$, along with the numerical results by
several other groups, violate the lower bound $\sim N^{1+\nu}/E$
suggested earlier in the literature \cite{kantor}. We also discussed
why this lower bound is incorrect and showed, based on conservation
of energy, that the correct lower bound for the pore-blockade time
for field-driven translocation is given by $\eta N^{2\nu}/E$, where
$\eta$ is the viscosity of the medium surrounding the polymer. Our
theoretical analysis has been supported by high precision  computer
simulation data, generated with a three-dimensional  self-avoiding
lattice polymer model.

Having worked out the physics of field-driven polymer translocation in
the absence of hydrodynamical interactions, it is worthwhile to
reflect on the scaling of pore-blockade times as a function of the
polymer length $N$ in the presence of hydrodynamical
interactions. Hydrodynamical interactions will modify the memory
kernel $\mu(t)$ --- changing it from
$t^{-(1+\nu)/(1+2\nu)}\exp(-t/\tau_{\tinytext{Rouse}})$ to
$t^{-(1+\nu)/(3\nu)}\exp(-t/\tau_{\tinytext{Zimm}})$
\cite{anom,anomlong}, where $\tau_{\tinytext{Zimm}}$ is the Zimm
relaxation time, scaling as $N^{3\nu}$ in good solvent for a polymer
of length $N$. This implies that the pore-blockade time will behave as
$N^{3\nu/(1+\nu)}$ under the influence of hydrodynamical
interactions. In this context we note that the scaling of the
pore-blockade time has been experimentally measured to scale as
$N^{1.26\pm0.07}$ \cite{storm}. In the scaling limit
$3\nu/(1+\nu)\simeq1.11$. The value for $\nu$ suggested in
Ref. \cite{storm} is $0.611\pm0.016$, for which
$3\nu/(1+\nu)\simeq1.14\pm0.02$, a bit closer to $1.26\pm0.07$. For a
physical explanation of the scaling of the pore-blockade times with
polymer length, the authors of Ref. \cite{storm} arrived at an answer
$2\nu$ using a macroscopic view of the translocating polymer, assuming
that the translational velocity of the centre-of-mass of the
untranslocated part is constant in time, and (implicitly) that the
memory kernel is a $\delta$-function. Our analysis in this paper, as
well as in Refs. \cite{anom,anomlong,forced,planar} based on memory
effects, therefore, casts serious doubts on the physical interpretation
of Ref. \cite{storm}: as we have repeatedly shown that the velocity of
translocation is not uniform in time, and the same part of the polymer
visits the pore a multitude number of times. Although so far our work
has not incorporated hydrodynamical interactions explicitly, it is
difficult to imagine that introducing hydrodynamical interactions will
mysteriously wipe out the entire memory effects in the polymer.

\section*{Acknowledgements}

Ample computer time on the Dutch national supercomputer facility SARA
is gratefully acknowledged.


\begin{thebibliography}{10}
\bibitem{noteunbiased} For unbiased translocation, $EN$ in this
argument is to be replaced by the difference in free energy (or
entropy) of a threaded polymer, corresponding to $s=N/2$, and the
translocated polymer, corresponding to $s=N$. This leads to the
inequality $\tau_d\geq\eta N^{1+2\nu}$.
\bibitem{note1} Note that with $\phi(t)$ a constant, strictly
  speaking, the integral (\ref{e7}) does not converge. The divergence
  stems from the assumption that $\mu(t)\sim t^{-(1+\nu)/(1+2\nu)}$
  holds all the way to $t\rightarrow 0$. This is clearly not true as
  can be seen from Fig. \ref{fig2}, which provides the required
  cutoff for the convergence of the integral (\ref{e7}).
\end{thebibliography}
\end{document}